\documentclass[prd,preprintnumbers,nofootinbib,aps,9pt]{revtex4}
\usepackage{subeqnarray}
\usepackage[title]{appendix}
\usepackage{epsfig,amsmath,amssymb,mathtools}
\usepackage{enumitem,setspace}
\usepackage{mathrsfs}
\usepackage{empheq}
\usepackage{soul}
\usepackage{centernot}
\usepackage[usenames,dvipsnames]{color}
\usepackage[pagebackref=true, colorlinks=true]{hyperref}
\definecolor{redish}{rgb}{0.7,0.2,0.0}  
\definecolor{bluish}{rgb}{0.2,0.5,0.8}
\hypersetup{linkcolor=redish,          
                  citecolor=blue,        
                  filecolor=magenta,      
                  urlcolor=bluish}          

\def \({\left(}
\def \){\right)}
\def \[{\left[}
\def \]{\right]}
\begin{document}
\title{Scientific value of the quantum tests of equivalence principle in light of Hilbert's sixth problem}
\author{Abhishek Majhi\footnote{A. M. is the corresponding author.}}
\email{abhishek.majhi@gmail.com}
\author{Gopal Sardar}
\email{gopalsardar.87@gmail.com}
\affiliation{Indian Statistical Institute,\\Plot No. 203, Barrackpore,  Trunk Road,\\ Baranagar, Kolkata 700108, West Bengal, India}

\begin{abstract}
{\color{black}In his sixth problem, Hilbert called for an axiomatic approach to theoretical physics with an aim to achieve precision and rigour in scientific reasoning, where logic and language (semantics) of physics play the pivotal role. It is from such a point of view, we investigate the scientific value of the  modern experiments to perform quantum tests of equivalence principle.} 
Determination of Planck constant involves the use of acceleration due to gravity of the earth $(g)$ that results in the force on a test mass. The equivalence between inertial mass and gravitational mass of a test object is assumed in the process of logically defining $g$ from the relevant hypotheses of physics. Consequently, if Planck constant is used as input in any experiment (or in the associated theory that founds such an experiment) that is designed to test the equivalence between inertial and gravitational mass, then it is equivalent to establish a scientific truth by implicitly assuming it i.e. a tautology. There are several notable examples which plague the frontiers of current scientific research which claim to make quantum test of equivalence principle. We question the scientific value of such experiments {\color{black} from Hilbert's axiomatic point of view. This work adds to the recently reported semantic obstacle in any axiomatic attempt to put ``quantum'' and ``gravity'' together, albeit with an experimental tint.}
\end{abstract}
\maketitle
\section{Introduction}
{\color{black}The significance of logic and language (semantics) in physics become evident  from the study of Hilbert's sixth problem, namely, {\it Mathematical Treatment of the Axioms of Physics}\cite{hilbertprob,corry2004} and the associated modern research that has germinated from such roots e.g. see ref.\cite{gorban} and the references therein (also see refs.\cite{gisin,gisin2,ll1,lpp}).  If we consider Einstein's views\cite{einphreal}, then ``{\it physics constitutes a logical system of thought}'' and ``{\it the justification (truth content) of the system rests in the proof of usefulness of the resulting theorems on the basis of sense experiences, where the relations of the latter to the former can only be comprehended intuitively.}'' While certainly the experimental observations (``{\it sense experiences}'') must have the final say on the essence of some theoretical construction, however, the inner consistency, of  the ``{\it logical system of thought}'' that underlies the theory, is necessary for the whole process of reasoning to be considered as scientific. Hilbert's call for axiomatization was aimed to achieve more precision and rigour in such reasoning. {\color{black} Such a call obviously is associated with a tension between exactness of logical truths and uncertain nature of experimental truths, the significance of which is manifested through Born's statement on p.81 of ref.\cite{born}: ``{\it a physical situation must be described by means of real numbers in such a way that the natural uncertainty in all observations is taken into account}''. While such tension has motivated lines of research investigation that concern the intuitive refinement of the language of physics itself\cite{ll1,lpp,chmajhi,nsg,gisin,gisin2}, but Hilbert's axiomatic point of view continues to retain its value in the mainstream literature\cite{gorban,corry2004,aqft1,aqft2,wightman,massgap} where the equations of physics are used to discuss physical phenomena and working principles of experiments with exact quantities devoid of the natural uncertainties involved in the experimental measurements. 
	} Recently, by adopting such a Hilbertian point of view, one of us (A. M.\cite{chmajhi}) has pointed out a semantic dilemma that one has to encounter in any attempt to axiomatically treat together ``quantum'' and ``gravity'' so as to pen down any theory of ``quantum gravity''\cite{carlip}. It is now our intent to investigate, from a similar point of view, the logic behind the modern experiments on the ``quantum'' tests of equivalence principle which are, nevertheless, very much motivated from the  ``quantum gravity'' mindset\cite{nature1,nature2,nature3,cqg1,prl1,prl2}.}


An experiment is considered to have a scientific value when (a) it verifies some hypothesis or theoretical prediction (b) {\color{black}it brings forth some newly observed phenomenon that is yet to be explained by theory}. However, an experiment of the type (a) can not depend on any input that explicitly or implicitly depends on the concerned hypothesis. If it does, then such an experiment verifies the truth by assuming it in the process. We consider this as a physical demonstration of a logical tautology. Such an experiment is devoid of any scientific value {\color{black}(unless it brings forth some hitherto unknown phenomenon that is not related to the concerned theory i.e. type (b))}. Here, we intend to discuss one such scenario, that involves a certain class of experiments (type (a)) which are designed to perform ``quantum'' tests of the equivalence principle \cite{nature1,nature2,nature3,cqg1,prl1,prl2}, {\it where and henceforth, by ``equivalence principle'' we  mean  ``equivalence between inertial mass and gravitational mass of an object'' \cite{bondi}}. These experiments inevitably involve the Planck constant (henceforth, to be denoted as ``$h$'') which justifies the word ``quantum''. {To be more particular, such experiments rely on atom interferometry where the De Broglie wavelength ($\lambda$) of the respective matter wave, with momentum $p$, needs to be calculated from the formula $\lambda=h/p$ and for this calculation $h$ is considered as a given input -- see refs. \cite{nature1,nature2,nature3,cqg1,prl1,prl2} and the relevant references there in.} 

We may express our concern in short as follows. Irrespective of the different measurement procedures to determine $h$ employed  till date, the corresponding theoretical analyses involve, either explicitly or implicitly, the use of $g$, where by the symbol ``$g$'', we  denote ``acceleration due to gravity of the earth'', or equivalently ``gravitational field of the earth''. For example, as we shall discuss here, in case of the Kibble balance method, $h$ is expressed in terms of $g$. On the other hand, in case of photo-electron emission  method, $g$ gets implicitly incorporated into the analyses through the processes by which ``charge of an electron'' is determined and used as an input.  However, $g$ can be {\it defined} from the axioms and hypotheses of physics, {\it if and only if the equivalence principle is assumed}. Therefore, the so called ``quantum tests of the equivalence principle''  are attempts to physically demonstrate a logical tautology and are devoid of any scientific value {\color{black} from Hilbert's axiomatic point of view. We shall point out that the assumption of the equivalence between ``inertial mass'' and ``gravitational mass'' is exact, and not associated with uncertainties, in the experiments that determine $h$. So, the ``quantum tests of the equivalence principle'' can not be interpreted as a recursive refinement of logic that verifies the equivalence principle with an uncertainty that is less than the uncertainty with which the equivalence principle is assumed to determine $h$. In fact, the determination of $h$ is based on the equivalence principle with zero or no uncertainty (i.e. exact) and then such $h$ is used as an input in the ``quantum tests of equivalence principle'' to verify the equivalence principle with non-zero uncertainty i.e. it is a case of adulteration (rather than refinement) of logic. Thus, if we take experimental uncertainty into account, the situation is even worse than a logical tautology. 
	
}

To present our arguments, we begin with a discussion regarding the definition of $g$, especially with a focus on how the assumption of equivalence principle is necessary for such a definition. Then, we discuss in what way the theoretical analyses for the different procedures of determining $h$ are founded on the assumption of the equivalence principle due to the involvement of $g$ and also due to the use of only one concept of ``mass'' regardless of any distinction as ``inertial'' and ``gravitational''. Finally, we conclude with a summary and some remarks regarding the status of this work in light of the new convention adopted by the science community in 2019\cite{bipm}. 


\section{Interpretations of the symbol ``$g$''}
There are two kinds of interpretations that we generally associate with the symbol ``$g$''. One is the {\it operational} interpretation and the other is the {\it logical} interpretation. 

We provide the operational interpretation of ``$g$'' in terms of what we can measure and it is the ``acceleration of a freely falling object {due to earth's gravity}'' which we determine experimentally (e.g. by dropping objects). {\color{black} To mention, the ``operational'' viewpoint of any definition in physics was advocated by Mach\cite{machmech}. We know that Einstein, who was highly influenced by Mach, took an ``operational'' approach to give meaning to ``time'' by stating our everyday experience of seeing the hands of a clock to mark the timing of an observed event like ``arrival of the train at the station''\cite{einsr1905}.  The operational perspective was revisited in an elaborate fashion by Bridgman\cite{logicphysics}. }

 However, if we restrict ourselves to being {\it completely} operational, we can not ask questions like the following -- ``how do we interpret that the acceleration is {\it due to earth's gravity}?'' This is because, in order to have a reasonable answer to this question, we need to consider hypotheses, axioms, etc. of physics, namely Newton's laws of gravity and motion, to provide the logical explanations corresponding to ``earth's gravity'' through mathematical expressions.  So, we need to be both operational (i.e. induce from experience) and logical (i.e. deduce from assumed truths i.e. axioms, hypotheses, etc.) -- {\color{black}we can not just stick to only one of the following stances -- ``a posteriori induction'', ``a priori deduction''. One can consult ref.\cite{quine} for a discussion regarding such issues.} We discuss, in what follows, that the equivalence principle is a {\it necessary} proposition for the logical (deductive) definition of ``$g$'' {\color{black}[see ref.\cite{tarskidef,tarskiundef} for some in depth discussion regarding such logical perspective of ``definition'']}.

 \subsection{Definition of ``$g$'' and the logical status of the equivalence principle}

At first sight it may appear to be an utter stupidity to have a discussion regarding the definition of ``$g$'' as it is ``too trivial''. Nevertheless, keeping in mind that the discussion is regarding some logical issues, we believe that the definition must be put into formal terms, rather than in a colloquial language, so as to clear any doubts regarding any of the individual propositions involved in the process, irrespective of its apparent triviality\cite{tarskidef}.  Therefore, in order to avoid any misjudgment by the concerned reader regarding the definition of ``$g$'', we consider formal statements, regardless of its simplicity and familiarity in the science community. It is much like what Hadamard has done to clear the doubt regarding the statement of  Huygens' principle in section (33) of ref.\cite{hadamard}. He has considered  ``simple formulae and statements'' in the form of propositions and has made a formal logical analysis in order to resolve doubts: ``{\it But, however simple the preceding formulae and statements, they have, nevertheless, opened somewhat important and lengthy scientific discussions, of which we have now to speak and which refer to what is called Huygens' Principle.}''


Apart from the above mentioned reasons,  we adopt such a path based on formal logic so as to put an emphasis on the necessity of the assumption of equivalence principle, as a formal statement, for the deductive definition of ``$g$''. In what follows, the symbols ``$\wedge,\iff,:=$'' denote ``logical conjunction, logical equivalence (if and only if), defined as'' respectively. {\color{black}In the context of writing down the propositions $A$ and $B$ citing ref.\cite{principia1,principia2} would have been relevant. However, those references do not contain the equations in the form that we use today. So, we have assumed that the reader is already acquainted with the Newton's laws.}
\begin{itemize}
	\item {\bf Proposition} $A$:  Newton's laws of motion apply to {\it an object whose motion is to be studied as a whole}, called {\it test object}. Associated concept of ``mass'' is called inertial mass ($m_I$). Acting force is given by 
	\begin{eqnarray}
		F=m_Ia,
	\end{eqnarray}
	where $a$ is the acceleration of the test object.
	\item {\bf Proposition} $B$: {\it The} test object obeys Newton's law of gravitation. Associated concept of ``mass'' is called gravitational mass ($m_G$). Force of gravitation on the  test object due to the earth is given by \begin{eqnarray}
		F=G\frac{m_GM_{earth}}{r^2}.
	\end{eqnarray} 
	Here, $G$ is a proportionality constant, $ M_{earth}$ is the gravitational mass {\color{black}(ignoring ``active/passive''  distinction, unlike in ref.\cite{bondi})} of the earth and $r$ is the distance between the test object and the earth, considering them as two points (point-masses). 
	\item {\bf Proposition} $C$: Equivalence principle holds:
	\begin{eqnarray}
		m_I=m_G,
	\end{eqnarray}
	where the meaning of the symbols ``$m_I$'' and ``$m_G$'' are explained by $A$ and $B$ respectively.\\
	
	{\bf Corollary:} $G$ is a fundamental constant if and only if $C$ is true for any type of material i.e. $m_{I(type)}=m_{G(type)}$ for any type of material like iron, aluminium, copper, etc. 
	

	\item {\bf Proposition} $D$: Concept of  ``gravitational field due to earth'', represented by the symbol ``$g$'', which is conceived through the observation of vertically falling objects towards the earth, is definable.
	\item {\bf Proposition} $E$: $g$ is definable {\it if and only if} Newton's laws of motion, Newton's law of gravitation and the equivalence principle hold simultaneously for the test object. Formally, $\left((A \wedge B)\wedge C\right)\iff D$.
	\\
	
	{\bf Corollary:}  $E$ implies the definition of $g$. Formally, 
	\begin{eqnarray}
		E\implies g:=\frac{GM_{earth}}{r^2}.
	\end{eqnarray}    
	\item {\bf Proposition} $F$: Considering $g$ as input, experiments are modeled to determine $h$. We call such models as  ``$g$-experiments'' i.e.
	\begin{eqnarray}
		g\text{-experiments}: (g\wedge\cdots).
	\end{eqnarray}
	The ``dots'' stand for propositions, other than $D$, which need to hold for the modeling. Such propositions can be both theoretical and experimental in nature. By ``experimental propositions'' we mean, the choices of appropriate physical conditions which are made by the experimenter in the laboratory. 
\end{itemize}

Let us explicate the relevance of the above propositions and the nature of the logical structure of the same as follows:
\begin{itemize}
	\item {\bf Step 1:} $A$ provides meaning to the symbol ``$m_I$''. $B$ provides meaning to the symbol ``$m_G$''. Hence, $C$ is {\it meaningless} without either of $A$ and $B$. Consequently, $C$ can neither be validated nor invalidated without $A$ and $B$ together. Also, validation or invalidation of $C$ does not affect the validity of $A$ or of $B$.  
	\item {\bf Step 2:} $D$ is {an assertion} about the possibility of {theoretically} defining the concept of ``gravitational field due to the earth'', represented by the symbol ``$g$''.  $E$ states the condition when such definition is possible by using $A, B$ and $C$. Thus, $E$ {\it implies} the definition of $g$. In other words, the definition of $g$ is implied by all the propositions from $A$ to $E$, albeit {\color{black}a} one-sided implication. 
	
	\item {\bf Step 3:} $F$ is the modeling of the $g$-experiments used to determine $h$. The meaning associated with the symbol ``$g$'' is explained by Step 1 and Step 2.  
\end{itemize} 
Therefore, we can draw the following conclusion:\vspace{0.1cm}


{\it Since $C$ is necessary for the definition of $g$, then $F$ is implied by $C$ i.e. the validity of the equivalence principle goes in as a necessary assumption when $g$ is used as an input to determine $h$. 
}

\subsection{``Completely operational'' or ``logical contradiction''?}
Although we have briefly argued just a bit earlier that we can not be completely operational, we would like to emphasize that point once again in light of the above logical analysis. One can certainly object to the conclusion (drawn in the earlier section) by arguing that, since $g$ is only implied by the successive propositions ($E\implies g$) and the reverse is not necessarily true (i.e. it is not an ``if and only if'' condition and rather a one-sided implication), then the tautology can be avoided. Instead of the above explained meaning of ``$g$'', a complete operational meaning can be assigned to ``$g$''as follows:\\

{\it The acceleration of a vertically falling object can be measured to have a value $9.81$ m/s$^2$ without referring to Newton's law of gravitation. Then, this explains the meaning of ``$g$'', and provides its numerical value that has been used in the $g$-experiments. In this way, there is no necessity of $B$ and hence, $C, D, E$ become meaningless. Only $A$ and the symbol ``$a$'' suffice and the introduction of the symbol  ``$g$'' is altogether unnecessary. We can have just ``$a$-experiments'' to determine $h$}.\\

However, in such a way of reasoning, the operational arguer {\it denies} the knowledge of the Newtonian\cite{principia1,principia2} (as well as Einsteinian\cite{einphil2}) theory of gravity that explains the phenomenon of falling object and consequently, denies the knowledge of the concept of ``gravitational mass''. So, if the arguer now claims to examine the validity of equivalence principle by using $h$, then he should {\it accept} the knowledge that he has already {\it denied}. Hence, the operational arguer runs into a logical contradiction, that is unacceptable.

At this point a diligent reader may be wondering whether our arguments imply that Eotvos et. al., in ref.\cite{eotvos1922}, have made a logical contradiction. The answer is certainly negative. It is true that the authors have considered $g$ in operational terms i.e. they used a directly measured value of $g$ in their analyses. However, their motive has been to judge how close are $g$ and $a$ if we {\it do not assume the equivalence principle} (e.g. see Box 1.2 on page 16 of ref.\cite{mtw}). Their case is different from the scenario where people consider the logical definition of $g$ by assuming the validity of equivalence principle and use the value of $g$ in operational terms if required to determine some quantity, which they use in turn to test the validity of equivalence principle. The issue will be more clear as we proceed and discuss what follows.

\section{Principles behind determination of $h$}
Now, let us discuss how the assumption of the equivalence principle is involved either explicitly or implicitly in the determination of $h$. Broadly, there are two categories of such experiments: (i) the Kibble balance experiments e.g. see refs.\cite{sanchez14, haddad17} and the relevant references therein {\color{black}(see ref.\cite{kibble} for Kibble's original work)} (ii) photo-electron emission experiments e.g. see refs.\cite{millikan16,huang20} and the relevant references in ref.\cite{huang20}.
 
\subsection{Kibble  balance experiments}   
In Kibble balance experiments \cite{sanchez14, haddad17}, a crucial step is the balancing between the gravitational force (due to earth) on a test mass $(m)${\color{black}(which includes the tare mass)} and force generated by a current $(I)$ carrying coil perpendicularly placed in a magnetic field. So, the relevant equation is\cite{sanchez14}:
 \begin{equation}\label{ph-1}
 F=mg=BLI
 \end{equation}
 where $L$ is the length of the coil and $B$ is the magnetic flux density.  This is called the weighing phase.

 The other phase, which is called the moving phase, is designed to measure $BL$ from the voltage developed  in the coil ($V$) when it passes with constant velocity $v$ through the magnetic field. The relevant equation is $V=BLv$,
 which is then used to eliminate $BL$ from eq.(\ref{ph-1}), under the assumption that $BL$ remains same in both phases, to obtain the Kibble equation:
 \begin{equation}
  mg=\frac{IV}{v}~,
  \end{equation} 

 Going through some further theoretical manipulations, corresponding to the measurement process (details can be found in ref.\cite{sanchez14}, unnecessary for the present purpose), the expression for $h$ comes out to be the following:
 \begin{eqnarray}
 	h=\gamma gvm\label{hexpression}
 \end{eqnarray}
 where $\gamma$ {\color{black}is a quantity with} physical dimension $[T]^2$ (time squared) and {\color{black}it depends on a collection of measured variables that depend upon the scaling parameters within the measurement technique.} 
 
 
  We note that in the theoretical analysis of the Kibble balance experiments, only {\it one} concept of ``mass'', denoted by ``$m$'', has been used and no categorization like ``inertial mass'' and ``gravitational mass'' has been made.  Furthermore, $g$  appears explicitly in the expression for $h$ i.e. in eq.(\ref{hexpression}). Therefore, the equivalence principle has been assumed {\color{black} and this assumption is exact because there is no mention of uncertainty regarding the equivalence of ``inertial mass'' and ``gravitational mass'' in refs.\cite{sanchez14,haddad17}.}
{\color{black} [{\it Remark}: Since Joule balance technique is just an {\color{black}alternative} of the Kibble balance and follows the same basic principle of balancing under gravity (see e.g. refs.\cite{hisplanck, joulebalance1, joulebalance2}), we do not discuss it separately. Our views regarding the Kibble balance that concerns the equivalence principle as the founding premise of the theoretical analysis, also applies for the Joule balance.]}

\subsection{Photo-electron emission experiments}  
 In  photo-electron emission  experiments,  as was first done by Millikan \cite{millikan16}, the theoretical analysis is founded upon the  photoelectric equation rooted to Einstein's work \cite{einphoto}: $h\nu=\phi+E_{max}$, where $E_{max}$ is the maximum kinetic energy of photo-electron, $\nu$ is the frequency of the incident light and  $\phi$ is the work function of the illuminated material. 
  
 It is found in such experiments that $E_{max}$ is proportional to $\nu$.  $E_{max}$ is determined by measuring the stopping potential $V$ from the equation $E_{max}=eV$, where $e$ is the charge of an electron. Therefore,  plotting $V$ along $y$-axis and $\nu$ along $x$-axis, $h/e$ is obtained as the slope of the straight line. Hence, {\it the value of  $h$ is determined from this slope $h/e$ by taking the value of $e$ as input.} 
 
 This fact does not change even in modern day experiments which involve photo-emission spectroscopy techniques to increase the precision of such experiments e.g. see ref.\cite{huang20} and the relevant references there in. So, it is worth understanding how  $e$ is determined (without using $h$ as input). We discuss the relevant methods in the following section.
 
{ \color{black}[{\it Remark:} There are examples of experiments, where experimenters claim to determine $e$ by using some value of $h$ as input in the process, without any elaborate discussion regarding how such value of $h$ is determined in the first place. For example, the experiments like those discussed in ref.\cite{jp18,w92}, which use the Single Electron Tunneling (SET) mechanism\cite{set1,set2}, use $h$ as input to determine $e$.  We keep any discussion regarding such experiments out of the present context because our motive is to discuss the methods of determining $h$ and not to assume its value as given.]}
 
 \subsection{Principles behind determination of $e$  (without using $h$ as input)} 
 There are broadly two methods of determining $e$ without using $h$ as input, which have been explored till date viz. oil-drop experiment and the x-ray spectroscopy method. While the former is explicitly dependent on the use of $g$, {\color{black}latter} is implicitly dependent on the use of $g$ through the determination of Avogadro constant in the process. Therefore, both the procedures are based on the assumption of equivalence principle, which we shall discuss in what follows. 
 \subsubsection{Oil-drop experiment}
 The oil-drop experiment is due to Millikan \cite{millikan11,millikan13} and Fletcher \cite{fletcher}.  The theoretical analysis behind the experiment can be traced back to ref.\cite{millikan11}, where the first equation has been written in the following way: ``{\it The relations between the apparent mass ${\bf m}$ of a drop, the charge $e_n$, which it carries, its speed, $v_1$ under gravity, and its speed $v_2$ under the influence of an electrical field of strength $E$, are given by the
	simple equation}
\begin{eqnarray}
	\frac{v_1}{v_2}=\frac{{\bf m}g}{e_nE-{\bf m}g}\quad\text{or}\quad e_n=\frac{{\bf m}g}{E}\(\frac{v_1+v_2}{v_1}\).''
\end{eqnarray}
The following clarification has been provided in a footnote: ``{\it The term `apparent mass' has been used to denote the difference between the actual mass and the buoyancy of the air.}''. Therefore, only {\it one} concept of ``mass'' (called ``actual mass'') has been used in such an analysis. No distinction such as ``inertial mass'' and ``gravitational mass'' has been made. Furthermore, the involvement of $g$ in the analysis is explicitly manifest from the expression for $e_n$, where $e_n$ stands for some integral multiple of $e$. Therefore, the equivalence principle has been assumed {\color{black} and this assumption is exact because there is no mention of uncertainty regarding the equivalence of ``inertial mass'' and ``gravitational mass'' in refs.\cite{millikan11,millikan13}.}

\subsubsection{X-ray spectroscopy method and the principles behind determination of Avogadro constant}
  The second method of determining $e$ relies on the use of x-ray spectroscopy to study crystal lattices as grating e.g. see refs. \cite{robinson35, robinson37} and the relevant references therein. This  method of determining $e$ requires the use of Avogadro constant $N_A$ as input \cite{hisplanck}. And, the theoretical analyses, which underlie the experimental determination of $N_A$ by various methods, are invariably founded upon only {\it one} concept of ``mass'' along with explicit involvement of $g$ \cite{perrin,avonum74, avonum76}, as we explain in what follows. {\color{black} Of course, by ``various methods'' we mean only those methods of measurements which neither use the value of $h$ nor use the value of $e$ as input to determine $N_A$.}

  In ref.\cite{perrin}, the final expression for $N_A$, that appears in the section called ``{\it Precise Determination of Avogadro's Constant}'', explicitly depends on $g$:
  \begin{eqnarray}
2.303 (RT/N_A) \log_{10} (n_0/n) = (4/3) \pi a^3 g (\Delta-\delta) {\bf h}.\label{avogadro}
  \end{eqnarray}
  The significance of the other symbols in the equation can be found in ref.\cite{perrin} and the symbol ``${\bf h}$'' represent some length and {\it not} Planck constant. The above expression is rooted to the buoyancy related analysis that follow from the earlier sections of ref.\cite{perrin}. Such analysis is based on only {\it one} concept of ``mass'' which becomes apparent from only {\it one} concept of ``density''. {\color{black}{A distinction as ``inertial mass'' and ``gravitational mass'' would have led to  two different notions of densities viz. ``inertial density'' and ``gravitational density''.}}

  To mention, $\Delta,\delta$ are the densities of the granular material and the inter-granular liquid, respectively, which have been used to perform the experiment.  Further, $g$ is explicitly manifest in eq.(\ref{avogadro}). Hence, the equivalence principle has been assumed in the process {\color{black} and this assumption is exact because there is no mention of uncertainty regarding the equivalence of ``inertial mass'' and ``gravitational mass'' in the concerned references.}

  In ref.\cite{avonum74}, the authors declare (in the second paragraph), the following: ``{\it We have made, instead, readily available highly spherical steel artifacts as local and temporary ``standards'' of density. Their masses were determined in terms of the U.S. National Standard (kilogram replica number 20) by well understood procedures.}'' This ``well understood procedure'',  described in ref.\cite{massmass}, is based on only {\it one} concept of ``mass'' and the use of $g$. Therefore, the equivalence principle has been assumed in the process {\color{black} and this assumption is exact because there is no mention of uncertainty regarding the equivalence of ``inertial mass'' and ``gravitational mass'' in ref.\cite{massmass}.}
  
  In a nutshell,  the determination of $N_A$ depends on the measured value of the density $(\rho)$ of the associated crystal and  $\rho$ is measured through buoyancy related experiments\cite{massbuo, fujii01, fujii02, fujii06, waseda01, waseda04}. Any such experiment is based on Archimedes' principle \cite{archimedes} where $g$ plays the role in defining ``the weight of an object''. Also, due to the inputs of the mass measurements, $g$ gets involved in the process as well\cite{massmass}. This is because mass measurements are done through mass comparators which are just balancing instruments with a founding principle based on the use of $g$ and only {\it one} concept of ``mass'' \cite{fb1a,fb1b,fb2}. Importantly, such mass measurements using mass comparators are involved in any modern experiment\footnote{We consider only those experiments which do not use $h$ as input in the process of determining Avogadro constant. This is because we are investigating the methods by which $h$ is determined.} that intend to determine $N_A$ e.g. the modern x-ray crystal density method discussed in refs.\cite{kuramoto17,becker09} indeed relies on mass comparators for mass measurement of the silicon crystal which becomes clear from the relevant references therein and especially from refs.\cite{picard06, massforavo}.
  
 Therefore, we can conclude from the above discussion that the determination of Avogadro constant $(N_A)$ is based on the assumption of the equivalence principle {\color{black} and this assumption is exact because there is no mention of uncertainty regarding the equivalence of ``inertial mass'' and ``gravitational mass'' in the concerned references.}

 \section{Concluding Remarks}
 {Let us conclude by providing a concise account of what we have discussed, followed by some crucial remarks regarding the status of this work in light of the new convention adopted in 2019\cite{bipm}.} While we define the symbol ``$g$'' to bear the meaning ``acceleration due to gravity of the earth'', we assume in course of such a definition that the equivalence principle (equivalence between inertial and gravitational mass) holds. Any measurement procedure of Planck constant ($h$), that involves the use of $g$, either explicitly or implicitly, is therefore dependent on the assumption of equivalence principle. While the Kibble balance method  explicitly involves the use of $g$, the  photo-electron emission experiments implicitly involves the use of $g$ through the measurement of the charge of an electron $e$. This is because the determination of $e$ through oil-drop experiment explicitly depends on the use of $g$ and the method of x-ray spectroscopy, to determine $e$, implicitly depends on $g$ due to the determination of Avogadro constant $(N_A)$ by measuring the crystal density through buoyancy related experiments (which require ``weight of an object'' to be defined) and mass measurements through mass comparators (which is based on balancing mechanism involving $g$ in principle). In view of these, we conclude that the modern day experiments which claim to make quantum tests of equivalence principle, are just attempts to physically demonstrate a logical tautology i.e. in such experiments one assumes  the equivalence principle to test the equivalence principle. {\color{black} Importantly, this assumption is exact because there is no mention of uncertainty regarding the equivalence of ``inertial mass'' and ``gravitational mass'' in the references that are concerned with the determination of $h$ directly or indirectly. Therefore, use of $h$, that has been determined with the assumption of the equivalence principle with no uncertainty, to test the equivalence principle with non-zero uncertainty, actually presents a case of adulteration of logic (worse than tautology).}

 Now, an apparently legitimate objection may be raised against this work, if one considers the new convention adopted by the scientific community in the year 2019 \cite{bipm}, in the following way:\vspace{0.1cm}

 {\small  {\bf Objection:} {\it Since 2019 the Planck constant ($h$) has been considered as a defining constant, in terms of which (and other defining constants), the kilogram -- unit of mass -- is defined. The distinction between ``inertial'' mass and  ``gravitational'' mass does not enter such a definition. }}\vspace{0.1cm}
 
 While such an objection may appear to be legitimate at first, however it is fallible to simple counter reasoning as we demonstrate in what follows. It is true that $h$ has been considered a defining constant since 2019 and it can be found to have been  clearly stated in the Preface of ref.\cite{bipm}. So, the respective value (in some system of units like SI system) is chosen to be exact and the base units like kilogram are defined in terms of $h$ and other defining constants. However, such a definition of kilogram, as given in ref.\cite{bipm}, is {\it theoretical} and yet to be realized through experiment. In fact, this has been clearly mentioned, on page no. 131 of ref.\cite{bipm}, that:\vspace{0.1cm}
 
{\small ``{\it The present definition fixes the numerical value of $h$ exactly and the mass of the prototype has now to be determined by experiment.}''}\vspace{0.1cm}

{\color{black}In fact, in the Appendix 2 of ref.\cite{bipm}, the expected uncertainty in such experimental realization of ``kilogram'', with reference to the mass of the international prototype of the kilogram, has also been mentioned.}

 That, the objection raised about the redundancy of the equivalence between inertial mass and gravitational mass, is empty of any essence, can be understood if we pose the following question regarding the above promise that ``the mass of the prototype has now to be determined by experiment'':\vspace{0.1cm}
 
 {\it What type of mass of the prototype is going to be determined by experiment -- inertial or gravitational?}  \vspace{0.1cm}
 
 In this work we have discussed precisely this issue regarding mass determinations in general and we have pointed out that one always assumes the equivalence between inertial mass and gravitational mass while performing such mass measurements (or even discussing the principles) because only one concept of ``mass'' is generally stated without any hint of such distinction. {\color{black}Importantly, this assumption of equivalence is exact and devoid of any uncertainty, which is distinct from the uncertainty expected to be associated with the experimental realization of kilogram reported in the Appendix 2 of ref.\cite{bipm}.} However, such an assumption is not stated formally and rather kept unmentioned or remains hidden. 
 
 Further, we may point out that the chosen value of $h$ has been determined through years of experimental research and then only it has been possible to reach a general consensus to consider an exact value, based on which  the base units are defined in ref.\cite{bipm}. What we have discussed in the article is that, all such procedures to determine the value of $h$, which have been performed over the years, are based on the assumption of the equivalence between inertial mass and gravitational mass, explicitly or implicitly. Now, if we consider the current convention of choosing $h$ to be {\it exact}, then it is implied by such convention that inertial mass and gravitational mass are {\it exactly} equivalent {\color{black}(irrespective of the expected uncertainty in the experimental realization of ``kilogram'' mentioned in the Appendix 2 of ref.\cite{bipm})}. Therefore, in light of the 2019 convention, the quantum tests of equivalence between inertial mass and gravitational mass \cite{nature1,nature2,nature3,cqg1,prl1,prl2}, are physical demonstrations of {\color{black}adulteration of logic (worse than tautology) because one starts with an exact $h$ (possible if and only if an exact equivalence between inertial mass and gravitational mass is assumed) and uses it for quantum tests of the equivalence between inertial mass and gravitational mass where the equivalence is associated with some non-zero uncertainty.}

{\color{black}In light of Hilbert's sixth problem and therefore, taking into account the importance of logic and language in physics, recently it has been reported by one of us (A.M.\cite{chmajhi}), how one should face a semantic or logico-linguistic dilemma in course of axiomatizing any theory that aims to take into account ``{\color{black}quantum}'' and ``gravity'' in the same framework. The present work brings to light similar scientific queries, but now with an association of the experimental aspect. Noting the recent upsurge in an interest in logic and language (semantics) of physics \cite{gorban,chmajhi,ll1,lpp,gisin,gisin2}, the present work adds to such research endeavours but with an added bit of direct association with questions concerning the experimental aspect as far as the critical question of treating ``quantum'' and ``gravity'' in the same framework is concerned, and even includes the context of the most recent 2019 convention of redefining of units\cite{bipm}. In view of this we hope that this present work can provide a different and fresh insight, for the concerned scientific community, as far as the experimental and the axiomatic aspect of  ``quantum gravity'' is concerned and, in particular, may germinate the seeds of certain reasonable doubts concerning the scientific reasoning that underlies the modern quantum tests of equivalence between inertial mass and gravitational mass\cite{nature1,nature2,nature3,cqg1,prl1,prl2}.  }


\vspace{0.2cm}
 
 {\it Acknowledgment:} The authors thank R. Radhakrishnan for pointing out ref.\cite{hadamard}. The work was accomplished while G. S. was visiting The Indian Statistical Institute, Kolkata. A. M. is supported by the Department of Science and Technology of India through the INSPIRE Faculty Fellowship, Grant no.- IFA18- PH208.
 
 {\it Conflict of interest statement:} On behalf of all authors, the corresponding author (A. M.) states that there is no conflict of interest. 

{\it Declaration on competing interest:} The authors declare that there is no competing interest.

\end{document}